\documentclass[a4paper,fleqn,usenatbib]{mnras}
\usepackage{newtxtext,newtxmath}
\usepackage[T1]{fontenc}
\usepackage{ae,aecompl}

\usepackage{graphicx}	% Including figure files
\usepackage{amsmath}	% Advanced maths commands
\usepackage{amssymb}	% Extra maths symbols
\usepackage{xcolor}     % Only for internal purpose

\usepackage[load-configurations = abbreviations,load=accepted]{siunitx} % unites
\usepackage{xspace} % For space after newcommand
\newcommand\ergcms{\si{erg.cm^{-2}.s^{-1}}\xspace}%
\newcommand\xcms{\si{cm^{-2}.s^{-1}}\xspace}%
\newcommand\cmstev{\si{cm^{-2}.s^{-1}.TeV^{-1}}\xspace}%
\newcommand\g{\ensuremath{\gamma}}%
\newcommand\hess{H.E.S.S.\xspace}%
\newcommand\fermi{\emph{Fermi}\xspace}%
\newcommand\fermilat{\emph{Fermi}-LAT\xspace}%
\newcommand\src{1ES\,2322-409\xspace}%
\providecommand{\e}[1]{\ensuremath{\cdot 10^{#1}}}

\title[VHE $\gamma$-ray discovery and MWL study of \src]{VHE $\gamma$-ray discovery and multi-wavelength study of the blazar \src}

\author[H.E.S.S. Collaboration]
{\Large\parbox{\textwidth}{
H.E.S.S. Collaboration,
H.~Abdalla$^{1}$,
F.~Aharonian$^{2,3,4}$,
F.~Ait~Benkhali$^{2}$,
E.O.~Ang\"uner$^{5}$,
M.~Arakawa$^{6}$,
C.~Arcaro$^{1}$,
C.~Armand$^{7}$,
M.~Arrieta$^{8,}$\thanks{Corresponding authors, \href{mailto:contact.hess@hess-experiment.eu}{contact.hess@hess-experiment.eu}},
M.~Backes$^{9,1}$
% ...
}}

\date{Accepted 2018 September 29. Received 2018 September 28; in original form 2018 May 15}

\pubyear{2018}

\begin{document}
\label{firstpage}
\pagerange{\pageref{firstpage}--\pageref{lastpage}}
\maketitle

%$^{*}$Corresponding authors
\clearpage
\parbox{\textwidth}{
$^{1}$Centre for Space Research, North-West University, Potchefstroom 2520, South Africa\\ 
$^{2}$Max-Planck-Institut f\"ur Kernphysik, P.O. Box 103980, D 69029 Heidelberg, Germany\\ 
$^{3}$Dublin Institute for Advanced Studies, 31 Fitzwilliam Place, Dublin 2, Ireland\\ 
$^{4}$National Academy of Sciences of the Republic of Armenia,  Marshall Baghramian Avenue, 24, 0019 Yerevan, Republic of Armenia \\ 
$^{5}$Aix Marseille Universit\'e, CNRS/IN2P3, CPPM, Marseille, France\\ 
$^{6}$Department of Physics, Rikkyo University, 3-34-1 Nishi-Ikebukuro, Toshima-ku, Tokyo 171-8501, Japan\\ 
$^{7}$Laboratoire d'Annecy de Physique des Particules, Univ. Grenoble Alpes, Univ. Savoie Mont Blanc, CNRS, LAPP, 74000 Annecy, France\\ 
$^{8}$LUTH, Observatoire de Paris, PSL Research University, CNRS, Universit\'e Paris Diderot, 5 Place Jules Janssen, 92190 Meudon, France\\ 
$^{9}$University of Namibia, Department of Physics, Private Bag 13301, Windhoek, Namibia\\ 
$^{10}$Department of Physics and Electrical Engineering, Linnaeus University,  351 95 V\"axj\"o, Sweden\\ 
$^{11}$Institut f\"ur Theoretische Physik, Lehrstuhl IV: Weltraum und Astrophysik, Ruhr-Universit\"at Bochum, D 44780 Bochum, Germany\\ 
$^{12}$DESY, D-15738 Zeuthen, Germany\\ 
$^{13}$Institut f\"ur Astro- und Teilchenphysik, Leopold-Franzens-Universit\"at Innsbruck, A-6020 Innsbruck, Austria\\ 
$^{14}$School of Physical Sciences, University of Adelaide, Adelaide 5005, Australia\\ 
$^{15}$Sorbonne Universit\'e, Universit\'e Paris Diderot, Sorbonne Paris Cit\'e, CNRS/IN2P3, Laboratoire de Physique Nucl\'eaire et de Hautes Energies, LPNHE, 4 Place Jussieu, F-75252 Paris, France\\ 
$^{16}$Laboratoire Univers et Particules de Montpellier, Universit\'e Montpellier, CNRS/IN2P3,  CC 72, Place Eug\`ene Bataillon, F-34095 Montpellier Cedex 5, France\\ 
$^{17}$Universit\'e Bordeaux, CNRS/IN2P3, Centre d'\'Etudes Nucl\'eaires de Bordeaux Gradignan, 33175 Gradignan, France\\ 
$^{18}$IRFU, CEA, Universit\'e Paris-Saclay, F-91191 Gif-sur-Yvette, France\\ 
$^{19}$GRAPPA, Anton Pannekoek Institute for Astronomy, University of Amsterdam,  Science Park 904, 1098 XH Amsterdam, The Netherlands\\ 
$^{20}$Friedrich-Alexander-Universit\"at Erlangen-N\"urnberg, Erlangen Centre for Astroparticle Physics, Erwin-Rommel-Str. 1, D 91058 Erlangen, Germany\\ 
$^{21}$Astronomical Observatory, The University of Warsaw, Al. Ujazdowskie 4, 00-478 Warsaw, Poland\\ 
$^{22}$Institut f\"ur Astronomie und Astrophysik, Universit\"at T\"ubingen, Sand 1, D 72076 T\"ubingen, Germany\\ 
$^{23}$Laboratoire Leprince-Ringuet, Ecole Polytechnique, CNRS/IN2P3, F-91128 Palaiseau, France\\ 
$^{24}$Instytut Fizyki J\c{a}drowej PAN, ul. Radzikowskiego 152, 31-342 Krak{\'o}w, Poland\\ 
$^{25}$Funded by EU FP7 Marie Curie, grant agreement No. PIEF-GA-2012-332350\\ 
$^{26}$School of Physics, University of the Witwatersrand, 1 Jan Smuts Avenue, Braamfontein, Johannesburg, 2050 South Africa\\ 
$^{27}$Universit\"at Hamburg, Institut f\"ur Experimentalphysik, Luruper Chaussee 149, D 22761 Hamburg, Germany\\ 
$^{28}$APC, AstroParticule et Cosmologie, Universit\'{e} Paris Diderot, CNRS/IN2P3, CEA/Irfu, Observatoire de Paris, Sorbonne Paris Cit\'{e}, 10, rue Alice Domon et L\'{e}onie Duquet, 75205 Paris Cedex 13, France\\ 
$^{29}$Nicolaus Copernicus Astronomical Center, Polish Academy of Sciences, ul. Bartycka 18, 00-716 Warsaw, Poland\\ 
$^{30}$Institut f\"ur Physik und Astronomie, Universit\"at Potsdam,  Karl-Liebknecht-Strasse 24/25, D 14476 Potsdam, Germany\\ 
$^{31}$Landessternwarte, Universit\"at Heidelberg, K\"onigstuhl, D 69117 Heidelberg, Germany\\ 
$^{32}$Univ. Grenoble Alpes, CNRS, IPAG, F-38000 Grenoble, France\\ 
$^{33}$Institut f\"ur Physik, Humboldt-Universit\"at zu Berlin, Newtonstr. 15, D 12489 Berlin, Germany\\ 
$^{34}$Obserwatorium Astronomiczne, Uniwersytet Jagiello{\'n}ski, ul. Orla 171, 30-244 Krak{\'o}w, Poland\\ 
$^{35}$Centre for Astronomy, Faculty of Physics, Astronomy and Informatics, Nicolaus Copernicus University,  Grudziadzka 5, 87-100 Torun, Poland\\ 
$^{36}$Japan Aerospace Exploration Agency (JAXA), Institute of Space and Astronautical Science (ISAS), 3-1-1 Yoshinodai, Chuo-ku, Sagamihara, Kanagawa 229-8510,  Japan\\ 
$^{37}$Department of Physics, University of the Free State,  PO Box 339, Bloemfontein 9300, South Africa\\ 
$^{38}$Department of Physics and Astronomy, The University of Leicester, University Road, Leicester, LE1 7RH, United Kingdom\\ 
$^{39}$Heisenberg Fellow (DFG), ITA Universit\"at Heidelberg, Germany\\ 
$^{40}$Yerevan Physics Institute, 2 Alikhanian Brothers St., 375036 Yerevan, Armenia\\ 
$^{41}$Now at Instituto de F\'{i}sica de S\~{a}o Carlos, Universidade de S\~{a}o Paulo, Av. Trabalhador S\~{a}o-carlense, 400 - CEP 13566-590, S\~{a}o Carlos, SP, Brazil\\ 
$^{42}$Oskar Klein Centre, Department of Physics, Stockholm University, Albanova University Center, SE-10691 Stockholm, Sweden
}

\clearpage

\begin{abstract}
%This is a simple template for authors to write new MNRAS papers.
%The abstract should briefly describe the aims, methods, and main results of the paper.
%It should be a single paragraph not more than 250 words (200 words for Letters).
  %No references should appear in the abstract.

A hotspot at a position compatible with the BL\,Lac object \src was serendipitously detected with H.E.S.S.\@ during observations performed in 2004 and 2006 on the blazar PKS\,2316$-$423. Additional data on \src were taken in 2011 and 2012, leading to a total live-time of 22.3h.
Point-like very-high-energy (VHE; $E>\SI{100}{\GeV}$) \g-ray emission is detected from a source centred on the \src position, with an excess of 116.7 events at a significance of $6.0\sigma$. The average VHE \g-ray spectrum is well described with a power law with a photon index $\Gamma=3.40\pm0.66_{\text{stat}}\pm0.20_{\text{sys}}$ and an integral flux $\Phi(E>\SI{200}{GeV}) = (3.11\pm0.71_{\rm stat}\pm0.62_{\rm sys})\times10^{-12} \centi \metre^{-2} \second^{-1}$, which corresponds to 1.1$\%$ of the Crab nebula flux above $\SI{200}{GeV}$.
%Contemporaneous or simultaneous m
Multi-wavelength data obtained with \fermi LAT, \textit{Swift} XRT and UVOT, \textit{RXTE} PCA, ATOM, and additional data from \textit{WISE}, GROND and Catalina, are also used to characterise the broad-band non-thermal emission of \src.  The multi-wavelength behaviour indicates day-scale variability. \textit{Swift} UVOT and XRT data show strong variability at longer scales. A spectral energy distribution (SED) is built from contemporaneous observations obtained around a high state identified in \textit{Swift} data. A modelling of the SED is performed with a stationary homogeneous one-zone synchrotron-self-Compton (SSC) leptonic model. The redshift of the source being unknown, two plausible values were tested for the modelling. A systematic scan of the model parameters space is performed, resulting in a well-constrained combination of values providing a good description of the broad-band behaviour of \src.
\end{abstract}


%%%%%%%%%%%%%%%%%%%%%%%%%%%%%%%%%%%%%%%%%%%%%%%%%%

%%%%%%%%%%%%%%%%% BODY OF PAPER %%%%%%%%%%%%%%%%%%

\section{Introduction}
\label{sec:Intro}

The High Energy Stereoscopic System\footnote{\url{https://www.mpi-hd.mpg.de/hfm/HESS/}} (H.E.S.S.) offering
a large field of view (\ang{5}), is not only suitable to cover extended sources of very-high-energy (VHE; $E>\SI{100}{\GeV}$) \g-ray emission, but also well-suited for unexpected discoveries in large areas surrounding point-source targets.
During an observation campaign on the blazar PKS\,2316$-$423 \citep{HESS_UL_2008}, a hotspot was observed at the position of another blazar, \src. This led to additional \hess observations of \src.
It is the third such fortuitous discovery of an extragalactic object by
ground-based air Cherenkov telescopes, following the discovery of the radio galaxy IC~310 with the MAGIC
telescopes \citep{MAGIC_IC310} and the blazar 1ES\,1312$-$423 with H.E.S.S.\@ \citep{HESS_1312}.

The blazar \src belongs to the most numerous class of extragalactic sources detected at VHE, the High Synchrotron Peaked \citep[HSP; $\nu_{sync}>10^{15}$ Hz, see][]{Ackermann:2015aa}.
The properties of blazars are a consequence of the orientation of their jets, which are aligned along or close to the line of sight,
thus modifying by relativistic beaming the apparent luminosity and variability time-scales measured by an observer on Earth.
The spectral energy distribution (SED) of blazars extends from radio to \g-rays and shows a two-humped structure, with a low-frequency component peaking between the optical and X-rays and a high-frequency hump peaking in the \g-ray domain.
The redshift of BL\,Lac objects is often difficult to determine because of the weakness or the absence of emission lines in their optical spectra,
and the frequent dilution of host galaxy absorption lines by the non-thermal radiation emitted by the compact object.

Non-thermal emission of \src has been detected at various wavelengths, including radio \citep{SUMSS},
infrared \citep{skrutskie:2006, wright:2010}, optical \citep{Jones2009}, and X-rays \citep{Elvis1992,Bade1992,Schwope2000}.
The source is also detected by \emph{Fermi} LAT in the GeV regime, and is present in the general
\citep[$\mathrm{E}>\SI{100}{\MeV}$, see][]{2015ApJS..218...23A} and high-energy \citep[$\mathrm{E}>\SI{10}{\GeV}$, see][]{FERMI_3FHL}
point-source catalogues. \src has been classified as BL Lac due to its featureless optical spectrum \citep{Thomas1998}.
Based on the broadband indices $\alpha_{radio-optical}$ and $\alpha_{optical-X-rays}$,
the position of its synchrotron peak has been estimated to $10^{15.92}$ Hz, resulting in the
  classification of the source as an HSP \citep{Ackermann:2015aa}.
The redshift of \src is unknown. The value $z\sim0.174$ reported by \citet{Jones2009} should be considered with caution, as it is based
on a low signal-to-noise-ratio spectrum which shows weak evidence for absorption lines corresponding to this redshift,
namely a single line at $\sim$\SI{6900}{\angstrom}. Beyond the fact that this is not enough to indicate unambiguously the redshift of the source,
it possibly corresponds to residual telluric absorption.

This paper presents the discovery of VHE $\gamma$-ray emission from \src with the H.E.S.S.\@ telescopes (Section~\ref{sec:H.E.S.S.}).
It presents also the compilation of data over a large spectral domain from infrared to high-energy \g-rays
(Section~\ref{sec:MWLdata}), and the modelling of a spectral energy distribution based on a subset of
simultaneous or contemporaneous data (Section~\ref{sec:Interpretation}).
Conclusions are presented in Section~\ref{sec:Conclusion}.

\section{H.E.S.S. discovery and analysis}
\label{sec:H.E.S.S.}

H.E.S.S.\@ is an array of telescopes located in the Khomas Highland of Namibia that detects VHE \g-rays
via the imaging atmospheric Cherenkov technique \citep{Aharonian:2006aa}.
The first phase of the experiment, lasting from 2002 until 2012, consisted of four \SI{13}{\metre} diameter telescopes
placed on the corners of a square of side \SI{120}{\metre}.
Since 2012, \hess operates in its second phase with the addition of a fifth \SI{28}{\metre} diameter
telescope placed at the centre of the array, which lowers the energy
threshold and enhances the sensitivity of the array at low energy.
This study only uses data taken during the first phase of the experiment.

The source \src was not part of the initial H.E.S.S.\@ blazar program, as TeV blazar candidates were selected in the
  early 2000's on the basis of their radio and X-ray properties \citep{Costamante:2002aa},
  at a time when no radio measurement was available for \src.
A first data set considered for this work corresponds to 9.3\;hours taken in 2004
and 2006 in the search for \g-ray emission from the blazar PKS\,2316$-$423.
During these observations, \src was often located close to the edge of the H.E.S.S.\@ field of view,
with angular distances relative to its centre between \ang{1.4} and \ang{2.2}.
A second data set corresponds to observations carried out in 2011 and 2012 during a campaign
dedicated to \src, for a total of 16.8\;hours. The source was then located
at offsets between \ang{0.5} and \ang{0.9}.
All observations were carried out at zenith angles ranging from \ang{17} to \ang{30}.
After data-quality selection, the total live-time amounts to 22.3\;hours.

Data were analysed using an updated version of the boosted decision trees (BDT) approach
described in \citet{2011APh....34..858B}, based on the same event parameters but including improvements in the BDT training process
and the use of $\gamma$/background discrimination cuts optimised for different templates of sources \citep{Khelifi2015Icrc}.
The analysis was performed with the `loose cuts' configuration, which requires a signal of at least 40\,photo-electrons in each camera that saw the shower,
and the use of discrimination cuts optimized for the detection of faint and soft-spectrum sources.
The corresponding energy threshold for the present data set is \SI{200}{\GeV}.
The results presented below were cross-checked with an independent calibration, reconstruction and analysis chain \citep{Parsons:2014aa}.

\begin{figure}
  \centering
  \includegraphics[width=\hsize]{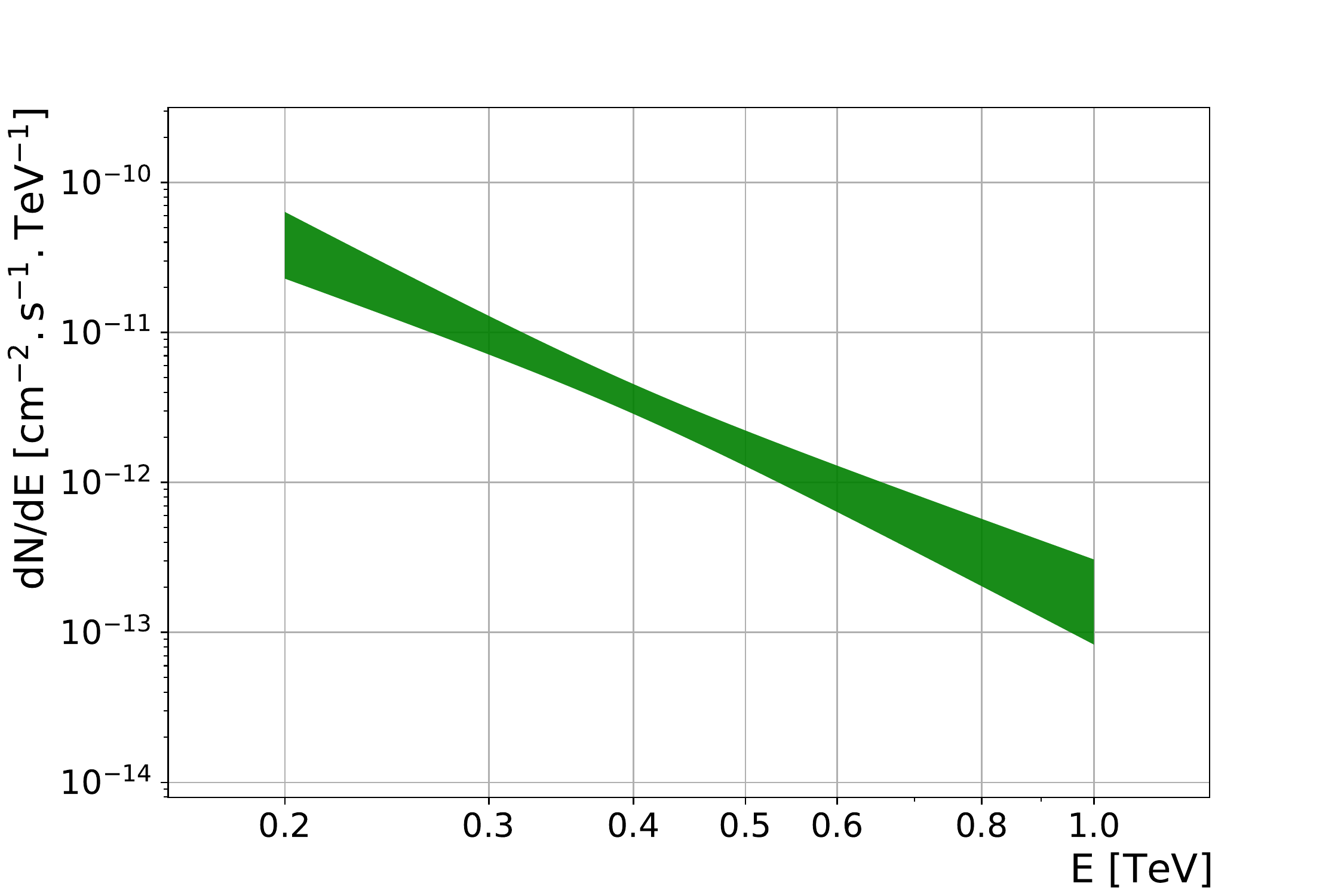}
  \caption{Time-averaged VHE spectrum of \src as a function of true energy.
    The green band correponds to the 68\% confidence-level provided by the maximum likelihood method for a power-law hypothesis.}
  \label{fig:Spec_HESS}
\end{figure}

\begin{figure}
  \centering
  \includegraphics[width=\hsize]{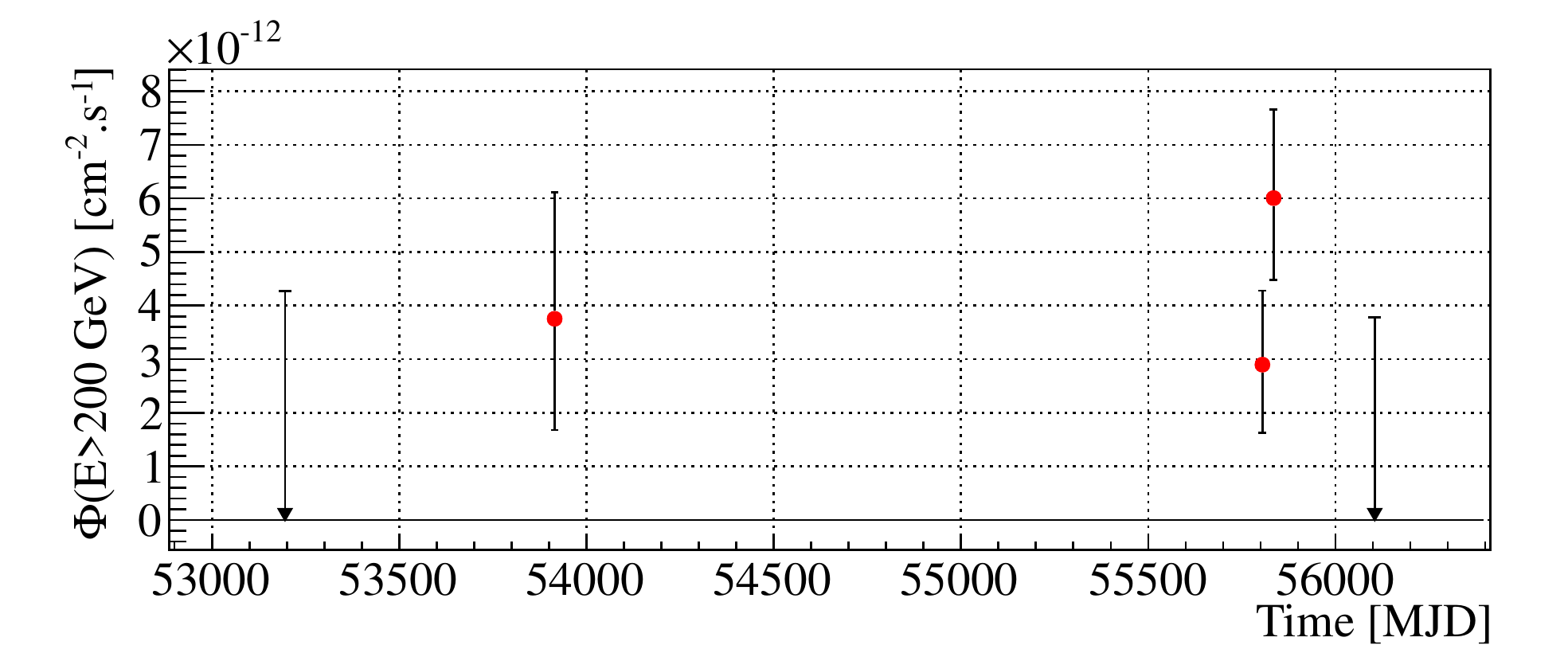}
  \caption{Monthly averaged integral fluxes of \src above 200\;GeV. Arrows correspond to 95\% upper-limits. Only statistical uncertainties are displayed.}
  \label{fig:LC_HESS}
\end{figure}

A $\gamma$-ray excess of 116.7 events with a statistical significance of $6.0\sigma$
\citep{Li:1983aa} was obtained within a
circular test region with a radius of \ang{0.11} centred on the 2MASS position of the source ($\alpha_{\text{J2000}}=23^{\text{h}}24^m44.68^{\text{s}}$, $\delta_{\text{J2000}}=-40\degree40\arcmin49.38\arcsec$).
Background was estimated using the reflected region method \citep{Berge:2007aa}.
A two-dimensional Gaussian fit of the excess, based on signal and background maps and the Point Spread Function (PSF) of the instrument,
yields a point-like source located at
$\alpha_\text{J2000}=23^{\text{h}}24^{\text{m}}48.0^{\text{s}}\pm4.8^{\text{s}}_{\text{stat}}\pm1.3^{\text{s}}_{\text{sys}}$
  and $\delta_\text{J2000}=-40\degree39\arcmin36.0\arcsec\pm1\arcmin12\arcsec_{\text{stat}}\pm20\arcsec_{\text{sys}}$.
This position is compatible with the 2MASS position of the source at the $\sim$1$\sigma$ level. No indication of extension was found.
The average differential photon spectrum of the source, shown in Fig.~\ref{fig:Spec_HESS}, was derived using a forward-folding technique \citep{Piron:2001aa}.
Considering a power-law hypothesis for the differential spectral shape, $\phi(E) = \phi_{0} (E/E_{\text{Ref}})^{-\Gamma}$,
where $E_{\text{Ref}} = \SI{0.40}{\TeV}$ is the decorrelation energy used as reference, and $\phi_{0}$ is the normalization at this energy, the spectral parameters are reconstructed as $\phi_{0} = (3.61\pm0.82_{\rm stat}\pm0.72_{\rm sys})\times10^{-12} \cmstev$ and $\Gamma=3.40\pm0.66_{\rm stat}\pm0.20_{\rm sys}$. The corresponding integral photon flux is
$\Phi(E>\SI{0.2}{TeV}) = (3.11\pm0.71_{\rm stat}\pm0.62_{\rm sys})\times10^{-12} \centi \metre^{-2} \second^{-1}$,
that is 1.1$\%$ of the Crab nebula flux \citep{Aharonian:2006aa} above the same threshold.
No statistically-significant evidence for spectral curvature or time variability was found.
The corresponding month-by-month light curve is shown in Fig.~\ref{fig:LC_HESS}.

\section{Multi-wavelength data}
\label{sec:MWLdata}

%To study the broad-band behaviour of the source, data were compiled over a larger spectral domain covering the H.E.S.S.\@ observations.
To study the broad-band behaviour of the source, additional data were compiled over different periods and over a large spectral domain.
These data, presented below, were taken from observations with \emph{Fermi} LAT (\SI{100}{\MeV}--\SI{500}{\GeV}), \textit{RXTE} PCA (2--\SI{60}{\keV}), \textit{Swift} XRT (0.2--\SI{10}{\keV}) and \textit{Swift} UVOT (170--\SI{650}{\nm}), GROND (Sloan optical $g'$, $r'$, $i'$ and $z'$, along with infrared $J, H \, \mathrm{and} \,  K$ filters), 2MASS (1.25, 1.65 and $\SI{2.17}{\micro \metre}$), \textit{WISE} ($3.6, 4.6, 12$ and $\SI{22}{\micro \metre}$), Catalina ($V$ band), ATOM (optical $B$ and $R$ filters), SUMSS ($\SI{843}{\MHz}$), GLEAM (80--\SI{300}{\MHz}), and TGSS (\SI{150}{\MHz}).
Only a fraction of the \emph{Fermi} LAT, \textit{Swift} UVOT, \textit{Swift} XRT, and GROND data are quasi-simultaneous.

% ...................................................................

\subsection{Fermi LAT}

The LAT instrument onboard the \fermi satellite detects \g-ray photons with energies between \SI{20}{MeV} and above \SI{300}{GeV}. Data were analysed using the publicly-available Science Tools \texttt{v10r0p5}\footnote{See \href{http://fermi.gsfc.nasa.gov/ssc/data/analysis/documentation}{http://fermi.gsfc.nasa.gov/ssc/data/analysis/documentation}.}. Photons in a circular region of interest (RoI) of radius \ang{10}, centred on the position of \src, were considered. The \texttt{PASS 8} instrument response functions (event class 128 and event type 3) corresponding to the \texttt{P8R2\_SOURCE\_V6} response were used together with a zenith-angle cut of \ang{90}. The model of the region of interest was based on the 3FGL catalogue \citep{2015ApJS..218...23A}. The Galactic diffuse emission has been modelled using the file \texttt{gll\_iem\_v06.fits} \citep{2016ApJS..223...26A} and the isotropic background using \texttt{iso\_P8R2\_SOURCE\_V6\_v06.txt}.

\fermilat data have been analysed for a period spanning from August 4, 2008 (MJD\,54682) to July 1, 2015 (MJD\,57204). Assuming a power-law spectral shape for \src, as per the 3FGL model \citep{2015ApJS..218...23A}, a binned likelihood analysis yields a detection with a Test Statistic $\text{TS}=787$ ($\sim$28$\sigma$) with an integrated photon flux of $F_\textrm{\SI{100}{\MeV}--\SI{500}{\GeV}} = \SI{7.17(97)e-9}{\xcms}$ and a photon index of $\Gamma=\SI{1.79(5)}{}$. The fit is performed iteratively, as described in \citet{HESS_PKS1510}. Using an alternative, more complex spectral model such as a log parabola does not significantly improve the fit. The most energetic photon detected from \src has an energy of $\sim$\SI{118}{\GeV} at a 95\% confidence level, as obtained using \texttt{gtsrcprob}.

Because the source PKS\,2325$-$408, whose brightness is comparable to the one of \src, is close-by in the RoI, at only \ang{0.69} from \src, and considering the large point spread function of the \emph{Fermi} LAT at low energies \citep[$\sim$\ang{5} at \SI{100}{\MeV}, $\sim$\ang{0.8} at \SI{1}{\GeV} and smaller than \ang{0.1} at \SI{500}{\GeV},][]{2013arXiv1303.3514A}, the data set was also analysed using a higher energy threshold of \SI{1}{\GeV}, in order to rule out leakage of photons from this nearby source. The corresponding results are compatible with the analysis performed using the full energy range, thus demonstrating that the modelling of the region of interest is under control in the entire energy range.

Data spanning from June 3, 2010 (MJD\,55350) to March 30, 2011 (MJD\,55650) will be used for the SED modelling (see Section~\ref{sec:Interpretation}, Fig.~\ref{LC_MWL}). In that time window, \src is detected with a TS of 198 ($\sim$14$\sigma$), with an integrated photon flux $F_\textrm{\SI{100}{\MeV}--\SI{500}{\GeV}} = \SI{8.15(230)e-9}{\xcms}$ and a photon index of $\Gamma=\SI{1.69(11)}{}$, and thus fully compatible with the entire data set within statistical errors.

\begin{figure}
  \centering
  \includegraphics[width=\hsize]{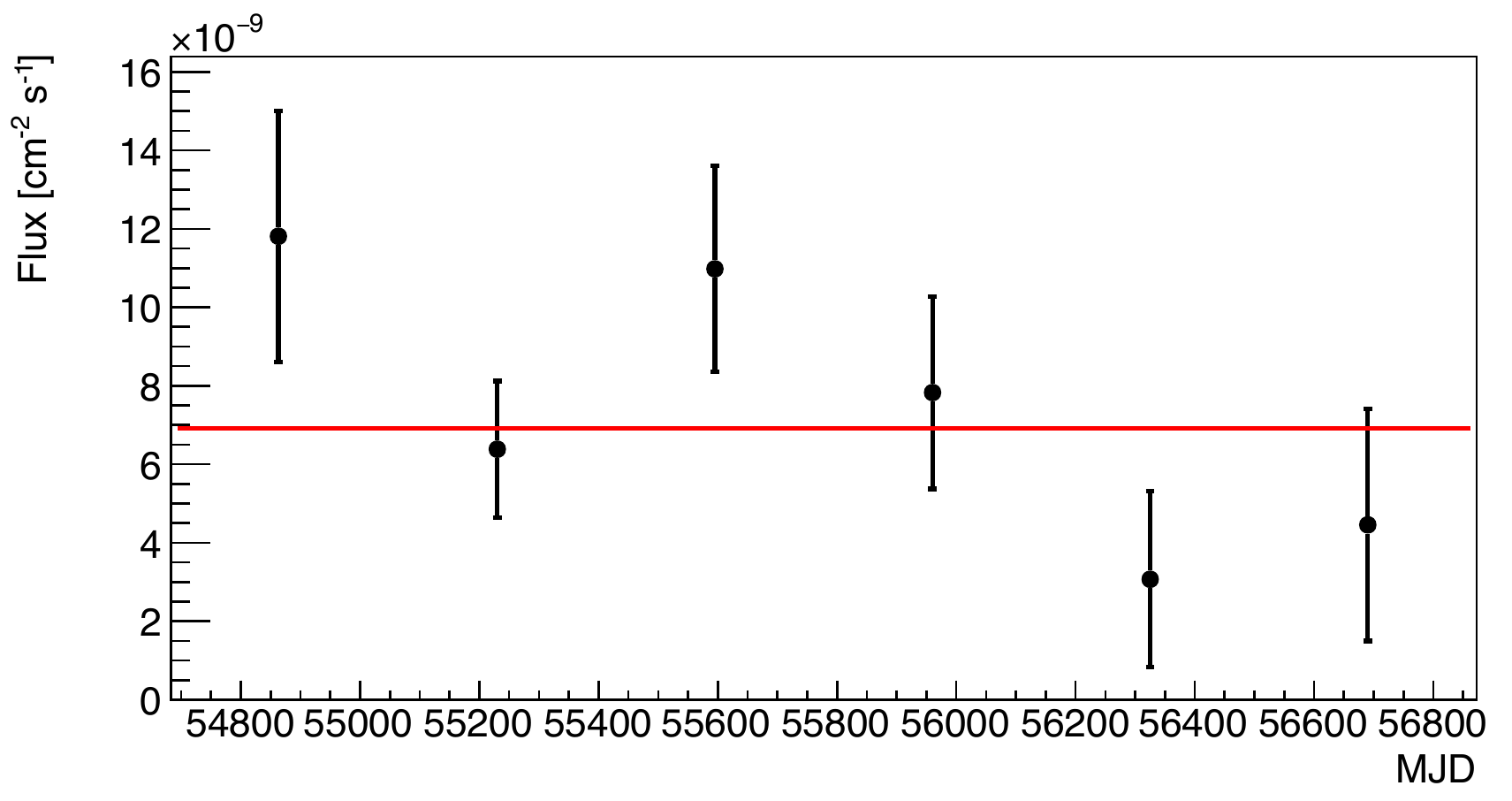}
  \caption{Yearly-binned light curve of \fermilat data, in the energy range \SI{100}{\MeV}--\SI{500}{\GeV}. The red line corresponds to a constant fit to the light curve.
  }
  \label{pic:fermilat_lc}
\end{figure}

The long-term variability pattern was tested using one-year time bins, with a constant fit to the light curve yielding $\chi^2/$n.d.f.$=8.58/5$ (see Fig.~\ref{pic:fermilat_lc}).
Even at shorter time scales, no variability is clearly seen in the monthly-binned light curve (see Fig.~\ref{LC_MWL}), a fit to a constant flux yielding $\chi^2/$n.d.f.$=79.8/67$ with a $p$-value of 0.14. This is consistent with the variability index of 41.32 reported by the \fermilat collaboration in the 3FGL catalogue \citep{2015ApJS..218...23A}.

\subsection{RXTE PCA}
\label{sec:RXTE}

X-ray observations of \src in the energy range 2--$\SI{60}{keV}$ were performed with the Proportional Counter Array \citep[PCA,][]{jahoda1996a} onboard the \textit{RXTE} spacecraft.
Seven pointings were taken nightly from the 15\textsuperscript{th} to the 21\textsuperscript{st} December 2011 for \src.
None of these pointings are contemporaneous with \hess observations.
The exposures of the PCA units are listed in Table~\ref{table:rxte}.
%The PCA is an array of five identical xenon-filled proportional counter units (PCUs) that cover the energy range 2--$\SI{60}{keV}$ with a total collecting area of $\SI{6500}{\cm\squared}$.
The analysis was performed using the standard
\texttt{HEASOFT} (v6.16) and \texttt{XSPEC} (v12.9) tools. The STANDARD2 data with a
time resolution of 16 seconds and with energy information in 128 channels were extracted
and filtered following the \textit{RXTE} Guest Observer Facility (GOF) recommended criteria.
Data were binned to ensure a minimum of 20 counts per bin.
Despite the broader energy range of the instrument, the source only presented sufficient statistics
in the 3--$\SI{7}{keV}$ energy range.
Thus, the average photon intrinsic spectrum from of all the 7 PCA observations was calculated in this energy range
for a power-law function. With the column density fixed at the Galactic value, i.e.
$N_{\text{H,tot}} = 1.67\e{20}\, \mathrm{cm}^{-2}$ \citep{Willingale:2013aa},
we obtained a photon index of $\Gamma = 2.80 \pm 0.15$ and a normalization at \SI{1}{\keV} of
 $\phi_0 = (7.26^{+1.80}_{-1.43})\e{-3} \, \mathrm{keV}^{-1}$ $\mathrm{s}^{-1}$ $\mathrm{cm}^{-2}$
(see Table~\ref{table:rxte_fit} for fit parameters).
The fit is not significantly improved considering a broken power-law shape, an F-test to compare the model fits yielding a probability of 0.043.

\begin{table*}
\centering                          % used for centering table
\begin{tabular}{c c c c c}        % centered columns (4 columns)
\hline\hline                 % inserts double horizontal lines
Date &  MJD & Our ID & \textit{RXTE} ID & Exp (ks)\\    % table heading
\hline                        % inserts single horizontal line
2011-12-15 02:11:44 & 55910.09 &  OBS A	& 96141-01-01-00 &  6.3\\
2011-12-16 01:38:56 & 55911.06 &  OBS B	& 96141-01-02-00 &  5.94\\
2011-12-16 23:32:48 & 55911.98 &  OBS C	& 96141-01-03-00 &  5.41\\
2011-12-18 00:33:52 & 55913.02 &  OBS D	& 96141-01-04-00 &  2.34\\
2011-12-19 01:34:40 & 55914.06 &  OBS E	& 96141-01-05-00 &  6.54\\
2011-12-19 21:54:40 & 55914.91 &  OBS F	& 96141-01-06-00 &  5.68\\
2011-12-21 00:29:36 & 55916.02 &  OBS G	& 96141-01-07-00 &  5.68\\
\hline                                   %inserts single line
\end{tabular}
\caption{Available \textit{RXTE} observations, corresponding dates and exposure times. }             % title of Table
\label{table:rxte}      % is used to refer this table in the text
\end{table*}

\begin{figure}
  \centering
\includegraphics[width=\hsize]{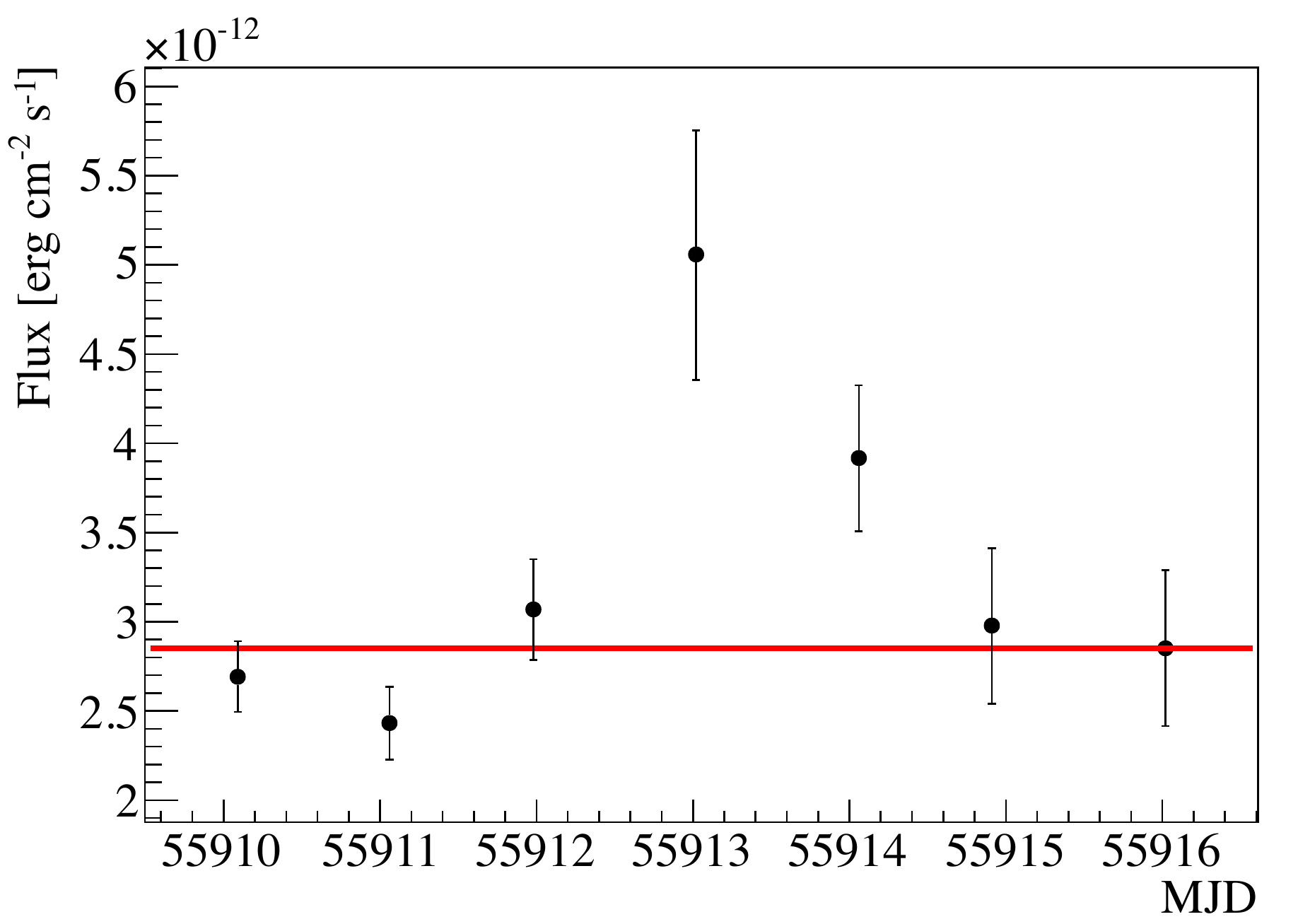}
\caption{\textit{RXTE} PCA light curve for all the available observations. Points correspond to the 3--$\SI{7}{keV}$ de-absorbed integrated energy flux. Dates are in MJD. The red line corresponds to a constant fit to the light curve.}
\label{pic:RXTE_diffobs}
\end{figure}

In order to obtain the integrated flux light curve, we performed observation-by-observation analyses. To do so,
because of low net count rates,  we fixed the photon index of all individual observations to that of the average state,
i.e.\@ $\Gamma = 2.80 \pm 0.15$. Corresponding fluxes in the
3--\SI{7}{\keV} range along with fit results are presented in Table~\ref{table:rxte_fit}. The source showed a small flare between the 18\textsuperscript{th}
and the 19\textsuperscript{th} December (Fig.~\ref{pic:RXTE_diffobs}). The fit of a constant to the light curve indicates evidence for variability with a
chance probability of $\sim$0.1$\%$.
%However, because of poor statistics, the average spectrum was considered for the multi-wavelength (MWL) SED of the source.

\begin{table*}
\centering                          % used for centering table
\begin{tabular}{c c c r c }        % centered columns (4 columns)
\hline\hline                 % inserts double horizontal lines
Our ID & Photon Index & Normalization at \SI{1}{\keV} & $\chi^2_{red}$ (d.o.f.) & F \\    % table heading
       &         & ($\rm keV^{-1} \, s^{-1} \,cm^{-2}$) &                        & (10$^{-12}$ $\rm erg \, cm^{-2}\, s^{-1}$) \\ %units
\hline
\hline
TOTAL  & 2.80 $_{-0.15}^{+0.15}$ & (7.26 $_{-1.43}^{+1.80}$)\e{-3} & 1.13 (7) & 2.93$_{-0.86}^{+0.82}$\\
\hline                    % inserts single horizontal line
OBS A	& 2.8 & (6.56$_{-0.49}^{+0.49}$)\e{-3} & 0.970 (8) & 2.69$_{-0.19}^{+0.20}$ \\
OBS B		& 2.8  & (5.94$_{-0.50}^{+0.50}$)\e{-3} & 0.334 (8) & 2.43$_{-0.20}^{+0.20}$\\
OBS C		& 2.8 & (7.50  $_{-0.68}^{+0.68}$)\e{-3} & 0.286 (8) & 3.06$_{-0.28}^{+0.28}$\\
OBS D	 &  2.8 & (1.23 $_{-0.17}^{+0.17}$)\e{-2} & 0.733 (8) & 5.05$_{-0.70}^{+0.70}$\\
OBS E		&  2.8 & (9.55 $_{-1.01}^{+1.01}$)\e{-3} & 0.561 (8) & 3.92$_{-0.41}^{+0.41}$ \\
OBS F		& 2.8 & (7.26 $_{-1.07}^{+1.07}$)\e{-3} & 0.192 (8) & 2.98$_{-0.44}^{+0.43}$\\
OBS G	 & 2.8 & 	(6.96 $_{-1.06}^{+1.06}$)\e{-3} & 0.831 (8) & 2.85$_{-0.43}^{+0.44}$ \\
\hline
\hline
                            %inserts single line
\end{tabular}
\caption{De-absorbed power-law parameters describing the differential photon flux obtained with XSPEC for \textit{RXTE} PCA observations (columns 2, 3 and 4), along with the 3--\SI{7}{\keV} de-absorbed integrated energy flux (column 5). See Section~\ref{sec:RXTE} for more details.}             % title of Table
\label{table:rxte_fit}      % is used to refer this table in the text
\end{table*}

\subsection{\textit{Swift} XRT and UVOT observations}
\label{sec:Swift}

\begin{table*}
\centering                          % used for centering table
\begin{tabular}{c c c c c}        % centered columns (4 columns)
\hline\hline                 % inserts double horizontal lines
Date & MJD & Our ID & \textit{Swift} ID & Exposure (ks)\\    % table heading
\hline                        % inserts single horizontal line
2009-11-17 13:37:00 & 55152.56 & OBS1 &  00031537001 & 4.43\\    % OBS1 in plotting
2010-03-30 06:53:00 & 55285.28 & OBS2 &  00040685001 & 1.18\\	 % OBS 4 in plotting
2010-03-30 08:33:00 & 55285.35 & OBS3 &  00040685002 & 4.50\\	% OBS 5 in plotting
2010-10-30 05:33:00 & 55499.23 & OBS4 &  00041657001 & 1.13\\   % High state!, OBS8 in plotting
2010-10-30 10:24:01 & 55499.43 & OBS5 &  00041656001 & 1.23\\	% OBS 7 in plotting
2012-11-04 02:51:00 & 56235.11 & OBS6 &  00040854001 & 1.24\\	%
2013-10-09 18:41:29 & 56574.77 & OBS7 &  00031537002 & 3.97\\
2013-10-12 00:54:00 & 56577.03 & OBS8 &  00031537005 & 3.12\\
\hline                                   %inserts single line
\end{tabular}
\caption{Available 8 \textit{Swift} observations, corresponding dates, IDs and exposure times in kiloseconds. }             % title of Table
\label{table:swift}      % is used to refer this table in the text
\end{table*}

X-ray and optical/UV observations of \src were performed with the XRT and UVOT detectors onboard the \textit{Swift} spacecraft \citep{burrows:2005aa}. The source was observed in 8 different occasions between November 2009 and October 2013 (OBS 1 to 8, see Table~\ref{table:swift}). None of these observations are contemporaneous with \hess data.
Results of the analysis of the datasets with the most comprehensive coverage both in XRT and UVOT energy bands are presented below, since they represent the best case scenario for modelling the SED of the source.
%For this reason, OBS3 and OBS7 are not considered for further analysis (see \ref{sec:UVOT} for details).
Note that OBS5 is omitted since the source is barely in the field of view and in a region with badly corrected exposure maps.

\subsubsection{XRT}
\label{sec:XRT}

X-ray observations of \src were performed with the XRT detector in Photon Counting (PC) mode in the 0.3--\SI{10}{\keV} energy range. Data were analysed following the standard \texttt{XRTPIPELINE} procedure\footnote{\url{http://www.swift.ac.uk/analysis/xrt/}} within the \texttt{HEASOFT} (v6.16) tools and were calibrated using the last update of CALDB. Source counts were extracted with the \texttt{xselect} tool from a circular region of radius 30 pixels ($\sim$ 71 arcsec), centred on the source, while background counts were extracted from a source-free region of radius 60 pixels. No pile-up correction was needed since the count rate was always lower than 0.5\,cts/s. The spectral analysis was performed via \texttt{XSPEC}, and data were binned to ensure a minimum of 20 counts per bin. The energy range was limited for each observation to ensure an acceptable number of event statistics. In the case of the observation with the best statistics, energies ranged from 0.4 to \SI{6.0}{\keV}, while for the integrated fluxes featured in the light curve (Fig.~\ref{LC_MWL}), a common range from 0.4 to \SI{4.0}{\keV} was selected. Following a procedure similar to that used for the \textit{RXTE} PCA data, a power law was fit to the different XRT data sets. Table~\ref{table:swift_fit} gathers the best-fit parameters derived for each individual observation considering a fixed Galactic column density (i.e.\@ $N_{\text{H,tot}} = 1.67\e{20}\, \mathrm{cm}^{-2}$), along with the integrated fluxes.

The XRT light curve (Fig.~\ref{LC_MWL}) illustrates variability but is not sufficiently well sampled to extract further information.
A closer look at the spectral index and flux values from Table~\ref{table:swift_fit} reveals a "harder when brighter" trend, with a correlation coefficient of -0.70. The observation with the brighest flux, OBS4, presents the hardest spectral index, $\Gamma = 2.14 \pm 0.09$, closely followed by observation with the second brighest flux, OBS1 with an index of  $\Gamma = 2.35_{-0.06}^{+0.05}$. The observation with the faintest flux, OBS6, has the largest spectral index, $\Gamma = 2.67 \pm 0.17$. This is also visible in Fig.~\ref{fig:swift_sed}. The modelling presented in Section~\ref{sec:Interpretation} will focus on the high state of the source as seen by \textit{Swift} in OBS4.

\begin{table*}

\centering                          % used for centering table
\begin{tabular}{c c c r c }        % centered columns (4 columns)
\hline\hline                 % inserts double horizontal lines
Our ID & Photon Index & Normalization at \SI{1}{\keV}&  $\chi_{red}^2$(d.o.f.) & F \\    % table heading  Date &  Swift ID &
       &         & ($\rm keV^{-1} \, s^{-1} \,cm^{-2}$) &                        & (10$^{-12}$ $\rm erg \, cm^{-2}\, s^{-1}$) \\ %units
\hline                        % inserts single horizontal line
OBS1 & 2.35 $_{-0.06}^{+0.05}$  & $(2.32 \pm 0.07)\e{-3}$ & 1.08(52) & 8.09$_{-0.46}^{+0.46}$ \\ %2009-11-17 13:37:00 & 00031537001 &
OBS2 & 2.30 $_{-0.15}^{+0.15}$  & $(1.13 \pm 0.09)\e{-3}$ & 0.76(8)  & 3.95$_{-0.34}^{+0.24}$\\% 2010-03-30 06:53:00 & 00040685001 &
OBS3 & 2.42 $_{-0.07}^{+0.06}$  & $(1.51 \pm 0.06)\e{-3}$ & 0.86(35) & 5.26$_{-0.40}^{+0.40}$\\ % 2010-03-30 08:33:00 & 00040685002 &
OBS4 & 2.14 $_{-0.09}^{+0.09}$  & $(2.59 \pm 0.13)\e{-3}$ & 1.10(21) & 9.27$_{-0.31}^{+0.31}$\\ % 2010-10-30 05:33:00 & 00041657001 &
OBS6 & 2.67 $_{-0.17}^{+0.17}$  & $(1.01 \pm 0.09)\e{-3}$ & 1.01(7)  & 3.47$_{-0.52}^{+0.52}$\\ %2012-11-04 02:51:00 & 00040854001 &
OBS7 & 2.36 $_{-0.07}^{+0.07}$  & $(1.27 \pm 0.05)\e{-3}$ & 1.30(25) & 4.57$_{-0.31}^{+0.33}$\\ % 2013-10-09 18:41:29 & 00031537002 &  GRISM TYPE!!! 9.99309463002592e-12 9.05732600898202e-12 1.09320094892763e-12
OBS8 & 2.43 $_{-0.18}^{+0.19}$  & $(1.17 \pm 0.06)\e{-3}$ & 1.10(15) & 4.06$_{-0.80}^{+0.81}$\\ % 2013-10-12 00:54:00 & 00031537005 &
\hline                                   %inserts single line
\end{tabular}
\caption{De-absorbed power-law parameters describing the differential photon flux obtained with XSPEC for the available XRT observations, along with the 0.4--\SI{4.0}{\keV} de-absorbed integrated energy flux featured in the light curve shown in Fig.~\ref{LC_MWL}. See Section~\ref{sec:XRT} for more details.}            % title of Table
\label{table:swift_fit}      % is used to refer this table in the text
\end{table*}

\subsubsection{UVOT}
\label{sec:UVOT}

\begin{figure}
  \centering
  \includegraphics[width=\hsize]{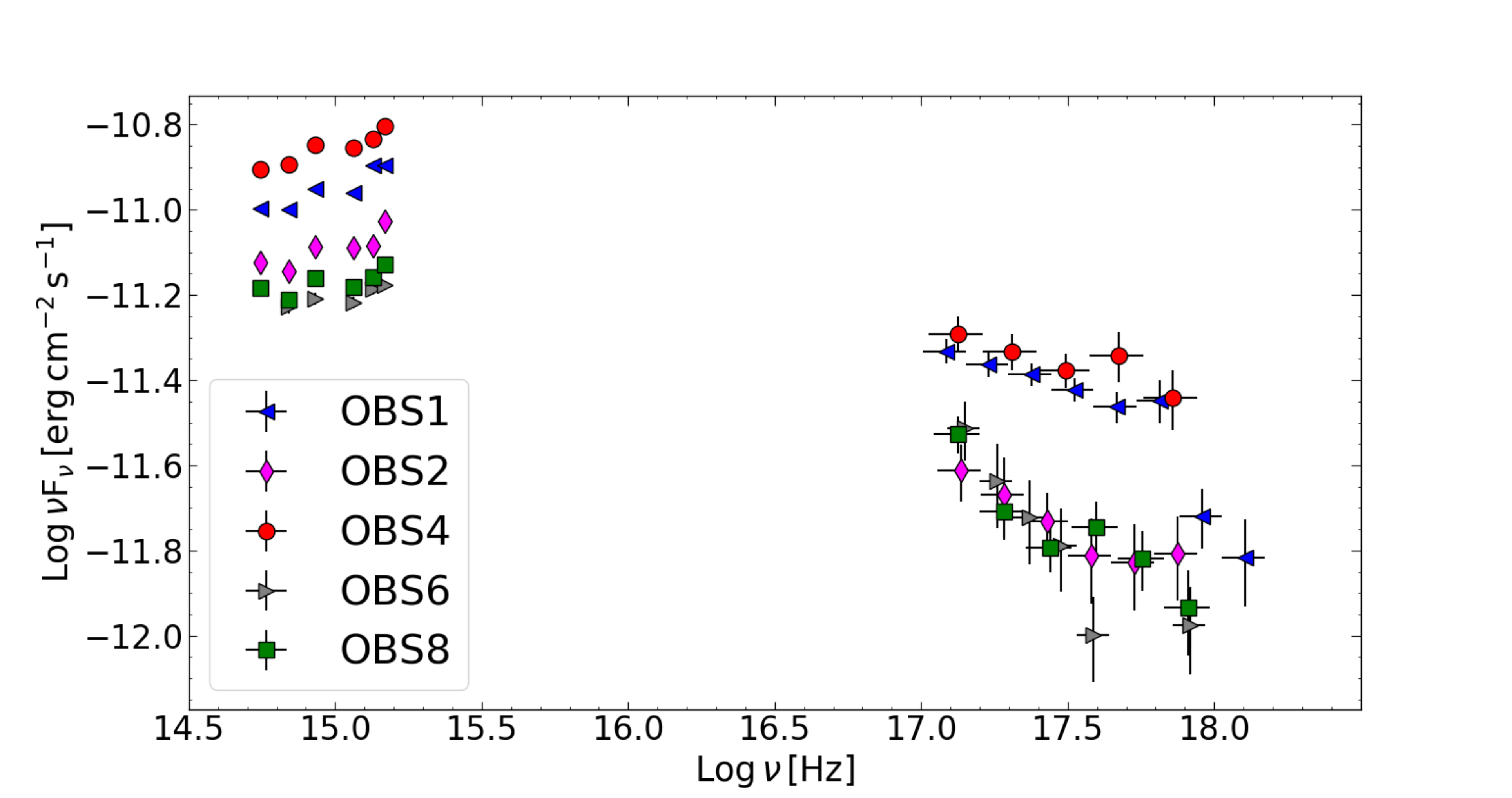}
  \caption{SED of different \textit{Swift} UVOT (absorption corrected) and XRT observations. Red points correspond to the highest state (OBS4) that will be afterwards used for the modelling of the source's energy distribution. Blue, magenta, red, gray and green points correspond to OBS 1, 2, 4, 6 and 8 respectively.
  }
  \label{fig:swift_sed}
\end{figure}

Simultaneous to XRT observations, the \textit{Swift} UVOT telescope \citep{roming:2005a} can acquire data in six filters: $v$, $b$ and $u$ in the optical band, $uw1$, $uvm2$ and $uvw2$ in the ultraviolet.
UVOT also features 2 grism modes, which provide rough spectroscopy in V and UV.
For OBS3, only the $uvw2$ filter was available, limiting its utility. For this reason OBS3 was omitted in our analysis. Likewise, OBS7 was of grism type precluding photometric analysis, and therefore was also omitted.
All available filters in each UVOT observation were searched for variability with the \texttt{UVOTMAGHIST} tool. Since no variability was observed in any filter, we then summed the multiple images within each filter. Source counts were extracted from a circular region of radius 5 arcsec centred on the source. Background counts were derived from an off-source region of radius 40 arcsec. Count-rates were then converted to fluxes using the standard photometric zero-points \citep{Poole:2008aa}. The reported fluxes are de-reddened for Galactic absorption following the procedure in \citet{roming:2009a}, with $E(B-V) = 0.0200 \pm 0.0004$.
The source exhibited variability between different observations, reaching a maximum flux around MJD\,55499 (see Table~\ref{table:swift_uvot1}
and Table~\ref{table:swift_uvot2} for UVOT exposure times and magnitudes for each passband, and Fig.~\ref{LC_MWL} for the corresponding light curve).
Fig.~\ref{fig:swift_sed} shows the resulting UVOT photometric points, along with the previously-mentioned XRT spectra.

\begin{table*}
\centering                          % used for centering table
\begin{tabular}{c | c |c c c| c c c| c c c}        % centered columns (4 columns)
\hline\hline                 % inserts double horizontal lines
Our ID &  Ext. & \multicolumn{3}{c|}{$uvv$ filter ($\lambda_0 = 5402\AA$)} & \multicolumn{3}{c|}{$uvb$ filter ($\lambda_0 = 4329\AA$)} & \multicolumn{3}{c}{$uvu$ filter ($\lambda_0 = 3501\AA$)}\\
       &  & Exp & Mag  & Flux  & Exp & Mag  & Flux & Exp & Mag  & Flux\\ % table heading \\
       &  & (s) & (Vega system)  & ($\rm mJy \, Hz^{-1}$)  & (s) &  (Vega system)  & ($\rm mJy \, Hz^{-1}$) & (s) &  (Vega system)  & ($\rm mJy \, Hz^{-1}$)\\ % table heading \\
\hline                        % inserts single horizontal line
OBS1 &  7 &  266.4 & 15.79$\pm$0.04  & 1.76$\pm$0.06 & 266.4 &16.16$\pm$0.03  & 1.39$\pm$0.04 & 266.4 &  15.17$\pm$0.03  & 1.23$\pm$0.03 \\
OBS2 &  1 & 95.2 & 16.09$\pm$0.07  & 1.33$\pm$0.08 & 95.2 & 16.50$\pm$0.05 &1.02$\pm$0.04 & 95.2 & 15.51$\pm$0.04  & 0.90$\pm$0.03 \\
OBS4 & 1  & 100.1 & 15.57$\pm$0.05  & 2.15$\pm$0.10 & 100.1 & 15.91$\pm$0.03 & 1.75$\pm$0.06 & 100.2 & 14.94$\pm$0.03  & 1.52$\pm$0.05\\
OBS6 & 2  & - & - & - & 40.2 & 16.62$\pm$0.07 & 0.91$\pm$0.06 & 40.1 & 15.83$\pm$0.04  & 0.67$\pm$0.04\\
OBS8 &  6 & 187.1 & 16.23$\pm$0.05  & 1.17$\pm$0.05 & 243.9 & 16.68$\pm$0.03  & 0.86$\pm$0.03 & 244.0 & 15.72$\pm$0.03  & 0.75$\pm$0.02\\
\hline                                   %inserts single line
\end{tabular}
\caption{Available UVOT photometric observations. The first column presents our observation ID, while the second states the number of individual images (extensions) within each observation. Exposure times, magnitudes and fluxes (non corrected for absorption) are given for different filters.}
\label{table:swift_uvot1}      % is used to refer this table in the text
\end{table*}

\begin{table*}
\centering                          % used for centering table
\begin{tabular}{c | c |c c c | c c c| c c c }        % centered columns (4 columns)
\hline\hline                 % inserts double horizontal lines
Our ID &  Ext. &  \multicolumn{3}{c|}{$uvw1$ filter  ($\lambda_0 = 2634\AA$)} & \multicolumn{3}{c|}{$uvm2$ filter ($\lambda_0 = 2231\AA$)}  & \multicolumn{3}{c}{$uvw2$ filter  ($\lambda_0 = 2030\AA$)}\\
&  & Exp & Mag  & Flux & Exp & Mag   & Flux & Exp & Mag   & Flux\\ % table heading \\
&  & (s) & (Vega system)  & ($\rm mJy \, Hz^{-1}$)  & (s) &  (Vega system)  & ($\rm mJy \, Hz^{-1}$) & (s) &  (Vega system)  & ($\rm mJy \, Hz^{-1}$)\\ % table heading \\
\hline                        % inserts single horizontal line
OBS1 &  7 & 534.2 & 15.04$\pm$0.03  & 0.86$\pm$0.02 & 555.6 & 14.96$\pm$0.03  & 0.80$\pm$0.02 & 1069.6 & 15.01$\pm$0.02  & 0.73$\pm$0.02\\
OBS2 &  1 & 190.7 & 15.37$\pm$0.04  & 0.64$\pm$0.02 & 289.8 & 15.29$\pm$0.04  & 0.59$\pm$0.02 & 381.7 & 15.34$\pm$0.03 & 0.54$\pm$0.01\\
OBS4 &  1 & 200.6  & 14.83$\pm$0.03  & 1.04$\pm$0.03 & 300.1 & 14.71$\pm$0.03  & 1.00$\pm$0.03 & 401.3 &  14.82$\pm$0.03  & 0.87$\pm$0.02\\
OBS6 &  2 & 85.6 & 15.75$\pm$0.06 & 0.45$\pm$0.02 & 120.8 & 15.50$\pm$0.06  & 0.48$\pm$0.03  & 161.2 & 15.80$\pm$0.05  & 0.35$\pm$0.02\\
OBS8 &  6 & 491.9 & 15.66$\pm$0.03  & 0.49$\pm$0.01 & 368.6 & 15.53$\pm$0.04  & 0.47$\pm$0.02 & 918.2 & 15.68$\pm$0.03  & 0.40$\pm$0.01\\
\hline                                   %inserts single line
\end{tabular}
\caption{Continuation of Table~\ref{table:swift_uvot1}. }             % title of Table
\label{table:swift_uvot2}      % is used to refer this table in the text
\end{table*}

\subsection{Optical and radio data}
\label{sec:archive}

\begin{figure*}
  \includegraphics[width = \hsize]{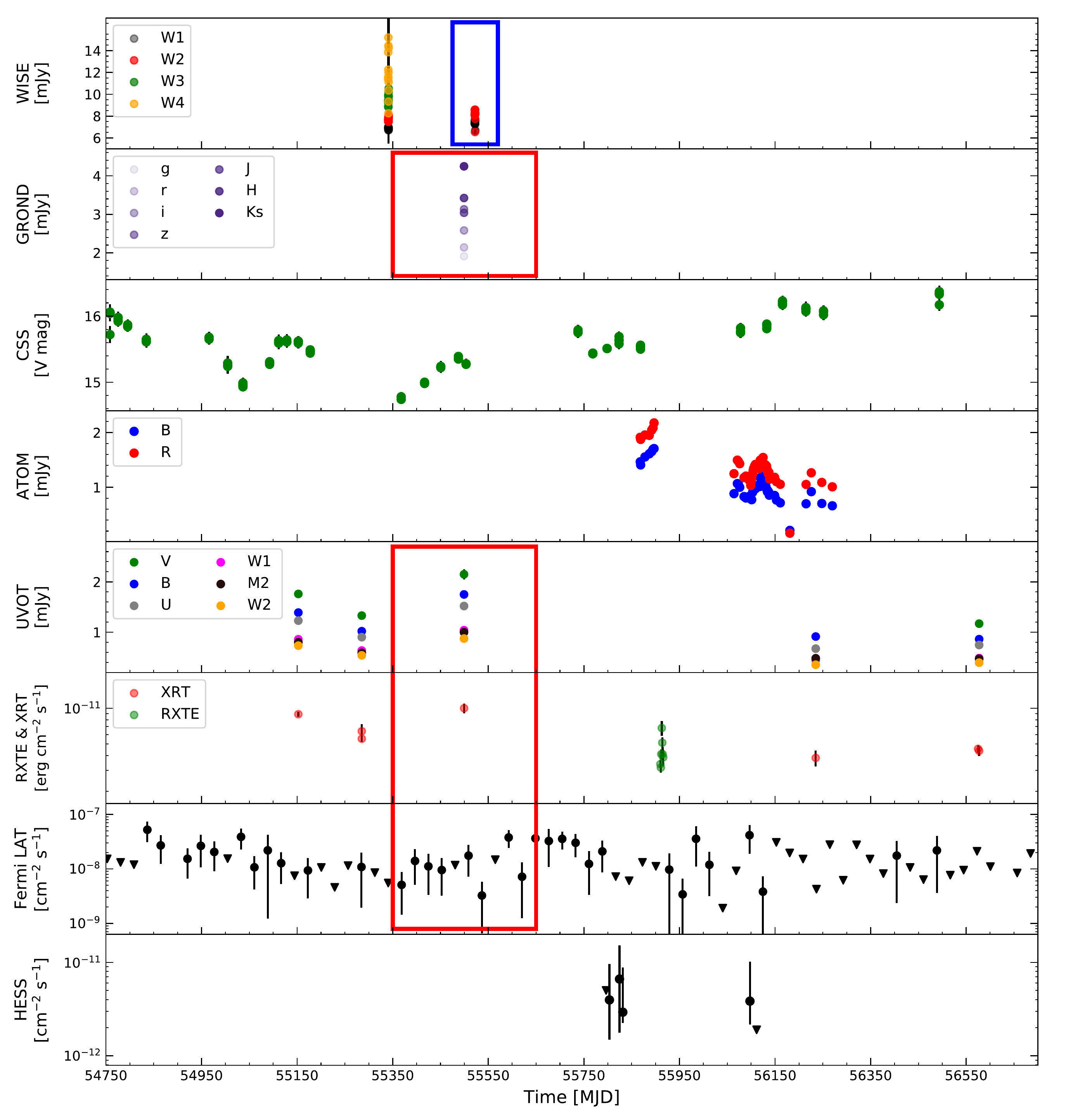}
  \caption{Multi-wavelength light curves of \src, from MJD\,54750 to MJD\,56700, in order of increasing energy, i.e.\@ \textit{WISE}, GROND, Catalina, ATOM, \textit{Swift} UVOT (non corrected for absorption), \textit{Swift} XRT, \textit{RXTE} PCA, \emph{Fermi} LAT and H.E.S.S.\@ (from top to bottom). The red rectangle encompasses the available quasi-simultaneous data, i.e.\@ GROND, \textit{Swift} UVOT and XRT, and \fermilat data, which are used for the modelling of the \textit{Swift} high state of the source,  while the blue rectangle shows the contemporaneous \textit{WISE} data, also considered for the modelling.
The GROND, Catalina, ATOM and \textit{WISE} light curves correspond to the data presented in Section~\ref{sec:archive}. The UVOT light curves are presented in Table~\ref{table:swift_uvot1} and Table~\ref{table:swift_uvot2}, the XRT light curve in Table~\ref{table:swift_fit} and the PCA light curve in Table~\ref{table:rxte_fit}. The \fermilat and H.E.S.S.\@ light curves correspond to 28 days-averaged and weekly-averaged fluxes, respectively (triangles correspond to 95\% upper-limits). Note that a zoom into the epoch for which most MWL observations were taken has been applied, so the totality of existing H.E.S.S.\@ data (which is considered for the modelling) is not shown in this picture (see Fig.~\ref{fig:LC_HESS} for the whole H.E.S.S. light curve).
Only statistical uncertainties are displayed. Note that the high-energy light curves are shown in logarithmic scale.}
  \label{LC_MWL}
\end{figure*}

The Automatic Telescope for Optical Monitoring (ATOM) is a 75 cm telescope located on the H.E.S.S.\@ site \citep{hauser:2004a}. Data for \src in the R and B bands are scattered between MJD 55850 and 56300 with a typical sampling frequency of 1 day, and are only simultaneous with H.E.S.S.\@ observations for a brief period of time. The flux points are included in the light curve in Fig.~\ref{LC_MWL}. The source went into a state of increasing flux with a hint of a flare peaking between MJD 55850 and MJD 56000. A second smaller flare was also observed around MJD 56100-56150. The observed variability time-scale is shorter than the ATOM data sampling.

Data are also available from the Catalina Sky Survey
\citep[CSS\footnote{\url{http://nesssi.cacr.caltech.edu/DataRelease/}},][]{Drake:2009aa}
over the same period of time as the \fermilat data sampled here. The CSS consists of seven years of photometry taken with the Catalina Schmidt Telescope located in Arizona (USA). Fig.~\ref{LC_MWL} presents V magnitude light curve for \src, which is found to be highly variable in the optical, as also observed with ATOM.

GROND
\citep[Gamma-Ray Optical/Near-infrared Detector,][]{grond:2008a}
is a 7-channel imager mounted at the MPG/ESO 2.2 m telescope in La Silla, Chile. Three infrared bands ($\text{J} = \SI{1.24}{\um}$,
$\text{H} = \SI{1.63}{\um}$, $\text{K}_\text{s} = \SI{2.19}{\um}$)
and the Sloan optical bands ($\text{g}^{\prime} = \SI{475}{\nm}$, $\text{r}^{\prime} = \SI{622}{\nm}$, $\text{i}^{\prime} = \SI{763}{\nm}$, and $\text{z}^{\prime} = \SI{905}{\nm}$)
are observed simultaneously, which is particularly interesting to analyse rapidly-variable sources such as blazars. The photometric data points for our analysis were taken from \citet{Rau:2012aa}.
A single observation was taken on 2010-10-31 at 23:54 (see Table~\ref{table:swift} and the light curve in Fig.~\ref{LC_MWL}).
Because the UVOT measurements often start earlier than ground-based measurements, some fine-tuning is required in order to compare both data sets. According to \citet{kruhler2011a}, the spectral overlap of UVOT and GROND can be used to correct the data between both instruments. The variability-correction factor $\Delta_{m_{GR \rightarrow UV}}=0.3$ from \citet{Rau:2012aa} based on the mentioned spectral overlap has been applied so that GROND points can be compared directly to the corresponding simultaneous UVOT points of OBS4.

\textit{WISE} \citep[Wide-field Infrared Survey Explorer,][]{wright:2010}
 observations were considered for the SED too.  \textit{WISE} is an infrared-wavelength astronomical space telescope launched in December 2009. With a 40-centimeter-diameter (16-inch) aperture, it was designed to continuously image broad stripes of sky at four infrared wavelengths ($\SI{3.4}{\um}$, $\SI{4.6}{\um}$, $\SI{12}{\um}$ and $\SI{22}{\um}$) as the satellite orbits the Earth. For \src, the observations were taken in two different time windows: the four filters were active during the first one ($\sim$ MJD\,55340), but only two (W1 and W2) for the second one ($\sim$ MJD\,55530), which is the period contemporaneous to the simultaneous GROND and \textit{Swift} observations (see Fig.~\ref{LC_MWL}). \textit{WISE} light curves were closely inspected in search of variability during this contemporaneous period. The lack of it allows us to consider the averaged spectral points for the SED.

 Although not used for the SED modelling of the source because they are out of the time-window considered in this work, radio data from the Sydney University Molonglo Sky Survey \citep[SUMSS,][]{SUMSS}\footnote{According to the ASDC, between 1997 and 2003}, the TIFR GMRT Sky Survey \citep[TGSS,][]{TGSS} and the GaLactic and Extragalactic All-sky MWA Survey \citep[GLEAM,][]{GLEAMA,GLEAMB} are also available for \src, and are depicted in Fig.~\ref{SSC_SED_ALL}. 
 
 Likewise, there are non-simultaneous Two Micron All Sky Survey \citep[2MASS,][]{skrutskie:2006} data, taken on August 8\textsuperscript{th} 1999.
 Although not used for modelling purposes, we decide to show these data in the MWL SED (Fig.~\ref{SSC_SED_ALL}) to have a broad picture of the source's spectrum, regardless of simultaneity constraints.

\section{Modelling and discussion}
\label{sec:Interpretation}

The rich MWL dataset gathered on \src allows us to perform for the first time a detailed study of its broad-band emission from infrared to VHE. A key step before performing the SED modelling is to carefully select the data in order to avoid variability effects that can bias the reconstruction of the source parameters. In the following we build a quasi-simultaneous SED of \src and interpret it in the framework of the standard SSC model for blazar emission, which has been successful in describing the emission from $\gamma$-ray HSP blazars. Given that the redshift of the source is unknown, we perform a study for the tentative value of $z=0.17$ provided by \citet{Jones2009}, and then test $z=0.06$, close to the redshift of several galaxies found around \src in shallow surveys\footnote{Note that the lack of deep redshift surveys around the source renders a quantification whether \src is part of a group of galaxies impossible.} \citep{Jones2009,Vettolani1998,Shectman1996,Ratcliffe1996}.

The light curves of \src at different wavelengths are presented in  Fig.~\ref{LC_MWL}.
There is no evidence of strong long-term variability observed at \g-ray energies. Day-scale and month-scale variability is seen both in optical and X-ray wavelengths. X-ray and optical/UV fluxes measured with XRT and UVOT are correlated, with a correlation coefficient higher than 0.9 whatever the UVOT filter considered.

The period around MJD\,55499 is considered for further analysis, as it is the only one with a quasi-simultaneous broad-band data set including IR, optical, X-rays and \g-rays. It corresponds to OBS4 in \textit{Swift} data, which is the highest state both in optical/UV and soft X-ray wavelengths (see Fig.~\ref{fig:swift_sed}). The simultaneous GROND observations for this period of time help constrain the synchrotron component. The contemporaneous \textit{WISE} observations ($\sim$ MJD\,55530) are also considered for the synchrotron peak constraints, whereas the high energy bump can be defined with a subset of the \fermilat data, corresponding to 300 days around MJD\,55499, and the whole H.E.S.S.\@ data set.

\begin{figure*}
  \centering
  \includegraphics[width=17cm]{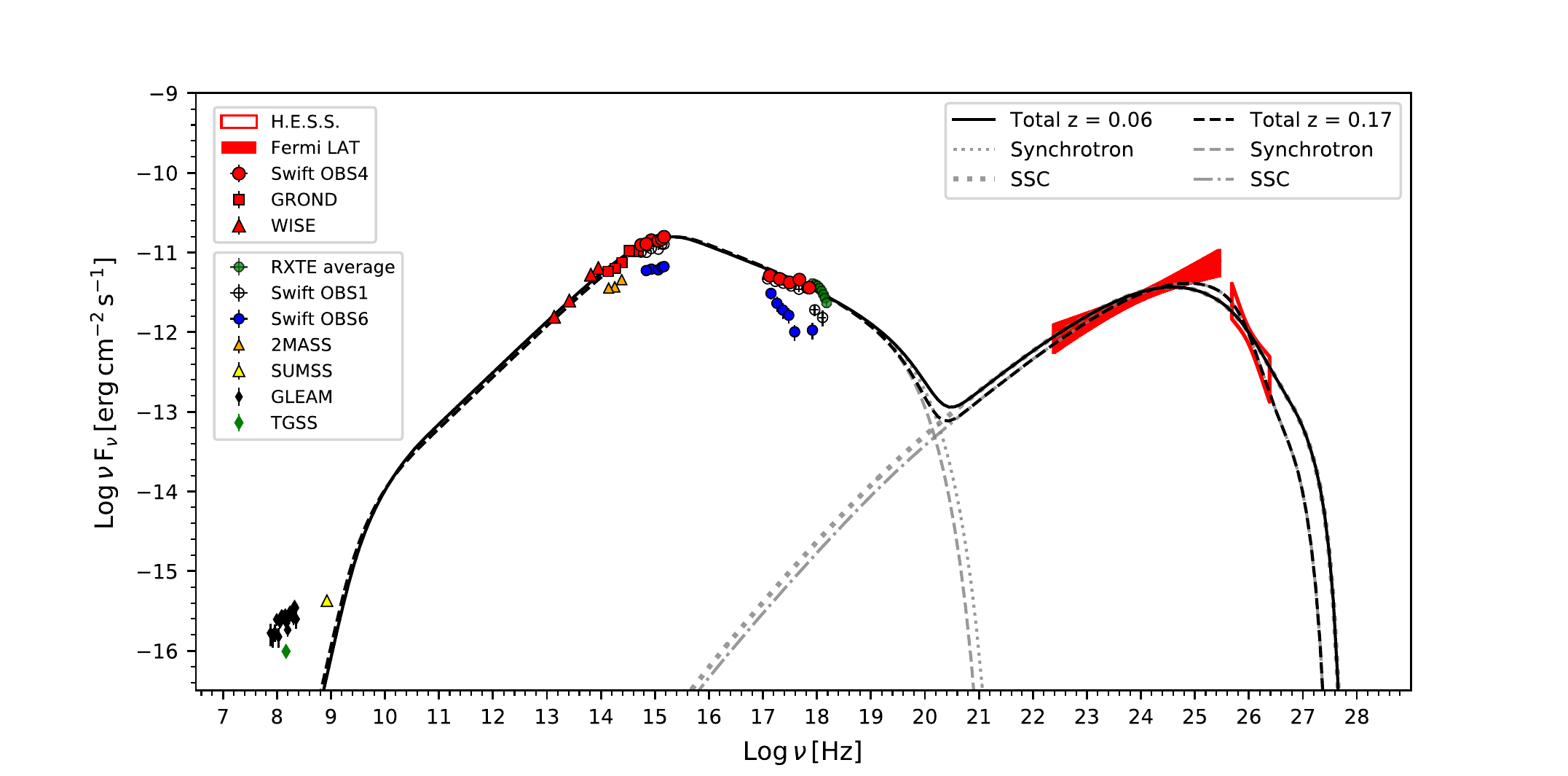}
  \caption{SSC modelling of the SED of \src considering two values for the redshift z = 0.17 and z = 0.06.
Red symbols correspond to data selected for the SED modelling (see Section~\ref{sec:Interpretation} for further explanation): the red hollow and filled bow ties represent the whole H.E.S.S.\@ and a sub-set of the \emph{Fermi}-LAT data, respectively; the red circles, squares and triangles represent the \textit{Swift} OBS4, GROND and \textit{WISE} data, respectively. Available data corresponding to other periods are also shown: green circles correspond to \textit{RXTE} data, black hollow and blue circles correspond to \textit{Swift} OBS1 and OBS6, respectively, orange triangles to 2MASS data, yellow triangles to SUMSS data, black diamonds to GLEAM data and green diamonds to TGSS data. Only statistical uncertainties are displayed. The dashed black line corresponds to a selected solution of the high redshift SSC model, whereas the solid black line corresponds to a selected solution of the low redshift SSC model. Absorption of the VHE emission by the extragalactic background light is accounted for following the model by \citet{2008A&A...487..837F}. See Table~\ref{table:modelparams} for input parameters for the model.
\label{SSC_SED_ALL}}
\end{figure*}

Since the available data set for \src (see Fig.~\ref{SSC_SED_ALL}) does not call for a more sophisticated approach, a one-zone stationary homogeneous synchrotron-self-Compton (SSC) model, based on \citet{Katarzynski:2001aa}, was chosen to provide a first characterisation of the parameters of the emission region. In this model, radiation is produced in a single zone of the jet approximated as a sphere of radius $R$, with a tangled magnetic field $B$, which moves through the relativistic jet at a small angle $\theta$ with respect to the line-of-sight. This description implies that the photons up to X-rays forming the first broad bump observed in the SED of BL\,Lac type blazars are produced by a population of relativistic electrons via synchrotron radiation. These synchrotron photons are then Inverse Compton (IC) scattered by the same population of electrons up to \g-ray energies, creating the second broad bump featured in the SED. The observed spectral shape requires a relativistic electron population that steepens with energy, which is conveniently modelled with a broken power-law (BPL) with a sharp high-energy break. This approach generally provides a good overall representation of the distribution of radiating particles.

The model can be completely described with 3 parameters related to the global features of the emitting region, namely the magnetic field $B$, the radius of the region $R$ and its bulk Doppler factor $\delta$, and with 6 parameters linked to the electron energy distribution, i.e.\@ the BPL indexes $n_1$ and $n_2$, the minimal and maximal electron energies $\gamma_{\rm min}$ and $\gamma_{\rm max}$, the break energy $\gamma_{b}$ and the normalization of the BPL $K$. 

Causality implies that the flux variability time-scale $t_{\rm var}$ is related to the size of the emitting region following $R \leq c t_{\rm var} \delta (1+z)^{-1}$, where $\delta$ is the Doppler factor and $z$ the redshift of the source.
For an estimate of this limit, we focused on modelling the flaring state of the source and applied the 1-2 day variability time-scale of the jet as seen in the X-ray band by \textit{RXTE} PCA.

\begin{figure}
  \centering
  \includegraphics[width=\hsize]{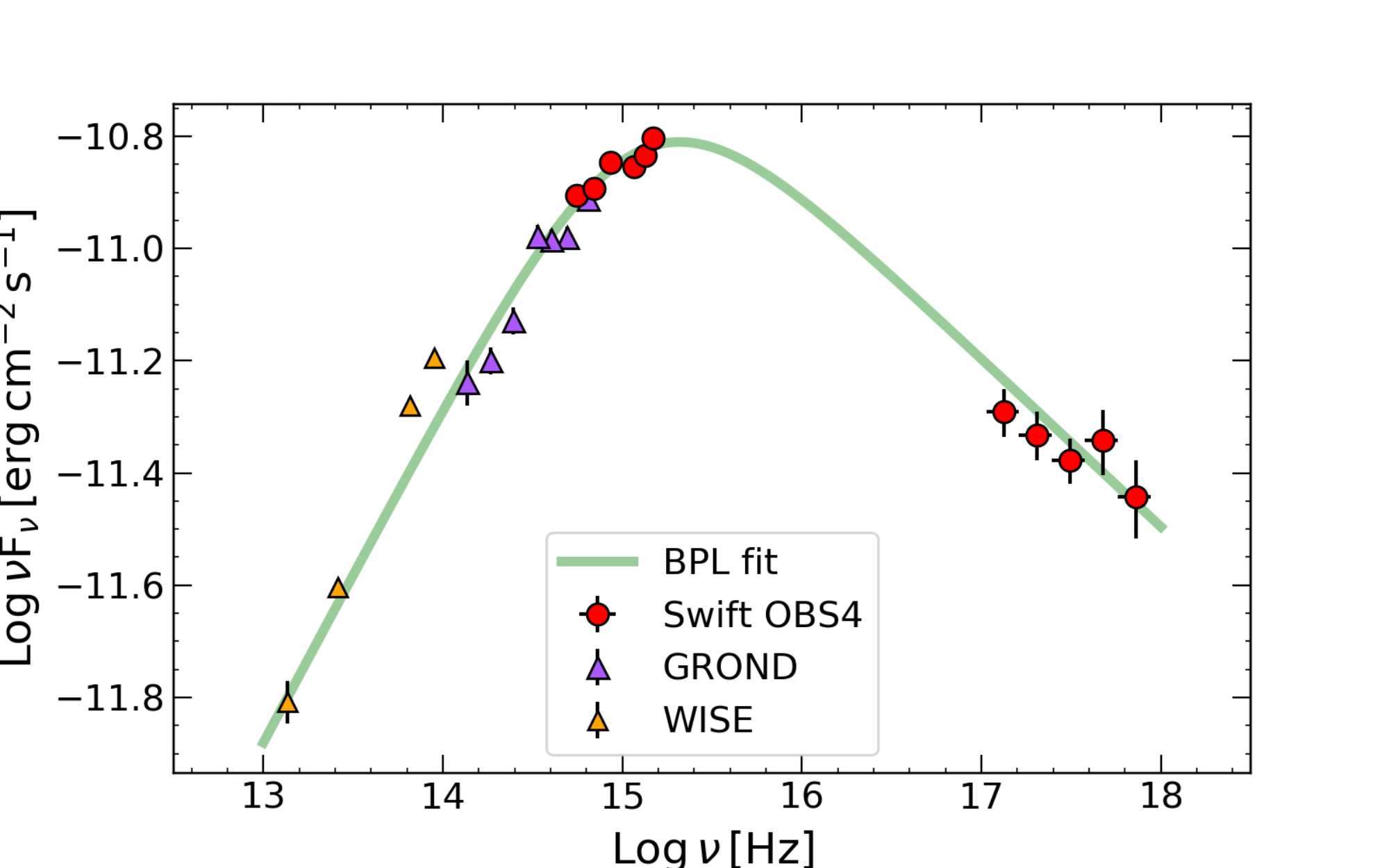}
  \caption{Spectral energy distribution of the source zoomed over the synchrotron-peak energy range. In green, the smoothly broken power-law fit of the high state XRT, UVOT and GROND simultaneous data along with contemporaneous \textit{WISE} data, which helps to determine the location of the synchrotron peak of the source.
  }
  \label{swift_high_logpar}
\end{figure}

From the broad-band SED in Fig.~\ref{SSC_SED_ALL}, one can see that both the synchrotron peak and the Compton peak are well determined by observational data. A precise determination of the synchrotron peak $\nu_{sync}$ and its luminosity $\nu f_{\nu, sync}$ is fundamental to constrain the SSC model. Considering that the synchrotron radiation from a BPL electron distribution is well described by a smoothly-broken power law, we fit this function to the selected SED data, i.e.\@ \textit{Swift} high state, GROND and \textit{WISE} data (see Fig.~\ref{swift_high_logpar}), to determine the position and luminosity of the synchrotron peak. The best-fit result is $\nu_{sync} = (2.09 \pm 0.12)\cdot 10^{15}$ Hz, and $\nu\ f_{\nu, sync} = (1.55 \pm 0.02)\cdot 10^{-11}$ erg cm$^{-2}$ s$^{-1}$. Since the SSC peak is located at approximately $\nu_{comp} \sim 10^{25}$ Hz, and the break energy $\gamma_{b}$ of the BPL electron distribution in the Thomson regime is expected to be located at $(3 \nu_{comp} / 4 \nu_{sync})^{1/2}$ \citep{tavecchio1998constraints}, we estimate that $\gamma_{b}$ is of the order of $10^{4}$.

Using the constraints on the peak positions and luminosities of the synchrotron and SSC components as starting points, the parameter space of the model was explored systematically using the algorithm developed by \citet{cerruti2013}. Five free parameters were investigated, in the following sub-set of the parameter space: $\delta\in[10, 100]$, $B\in[2,50]\, \mathrm{mG}$, $R\in[1\e{16}, 2\e{17}]\, \mathrm{cm}$, $\gamma_b\in[1\e{4},1\e{5}]$,  $K\in[1\e{-8}, 5\e{-6}]\, \mathrm{cm}^{-3}$, where $K$ is defined as the normalization of the electron distribution at $\gamma_b$.
Solutions outside the sampled parameter space do not exist, as they are excluded analytically following the approach described in \citet{tavecchio1998constraints}.
Two assumptions of the redshift of the source were probed, using the model by \citet{2008A&A...487..837F} to account for the absorption of the VHE emission by the extragalactic background light.
The indices of the particle distribution $n_1$ and $n_2$, being well constrained by the \fermilat and \textit{Swift} XRT spectra, respectively, were fixed to values within the uncertainties of the measured slopes that provided a maximum range of acceptable model solutions. This was necessary to reduce the large number of degrees of freedom of the model considered for the parameter scan. The values of $\gamma_{min}$ and $\gamma_{max}$ were also fixed, given the small impact of these parameters on the model SED. Taking into account the considerations above, $12^5$ SSC models were produced, computing for each of them the frequency and flux of the synchrotron peak, the flux and spectral index in the \hess energy band, and the flux and spectral index in the \fermilat energy band. To compare with the $\gamma$-ray observables, we calculate for every SSC model a fit with a power-law function over the \fermilat and \hess detection bands, obtaining the associated dependence of the spectral index and the flux at the instrumental decorrelation energy on each model parameter. In addition to these four $\gamma$-ray observables, the two synchrotron observables introduced above are also used in the following. Each of these six observables is then expressed as a function of the model parameters, producing a set of equations. With five variables, a system of five equations is enough to provide a unique constraint on each variable. We solve the system requiring that $log(\nu_{sync})\in[15.29, 15.34]$, $log(\nu f_{\nu, sync})\in[-10.81, -10.80]$, $log(\nu f_{\nu, LAT})\in[-11.64, -11.48]$, $log(\nu f_{\nu, \hess})\in[-12.185, -11.915]$, $\Gamma_{\hess}\in[-4.09, -2.71]$, where $\nu_{sync}$ is in Hz and $\nu f_\nu$ is in \ergcms. We further select solutions which satisfy the conditions on the LAT index $\Gamma_{LAT}\in[-1.89, -1.49]$ and the variability time-scale $t_{\rm var} < 1.5\, \mathrm{days}$. We also exclude solutions with $\delta > 100$, which are outside the explored parameter space and much higher than estimations from radio observations. Please note that the $\gamma$-ray observables include the systematic uncertainties, summed in quadrature to the statistical ones. Note also that no $\chi^2$ minimization is performed, as the algorithm simply selects SSC solutions which are compatible with the observations, as a numerical generalization of the \citet{tavecchio1998constraints} approach.
The values of the SSC solutions are provided in Table~\ref{table:modelparams}. For each solution the energy budget of the emitting region is calculated, and we provide the range of derived $u_e/u_B$ and $L$ values, where $u_e$ and $u_B$ are the kinetic and magnetic energy densities in the source frame and $L$ is the jet power.

Given the degeneracy of the SSC model and the correlations between different parameters, an optimal solution cannot be identified, but it is instructive to discuss selected examples. The selected solutions have bulk Doppler factors $\delta$ and energy density ratios $u_e/u_B$ close to the lower limits found from the parameter scans.

For the source redshift $z = 0.17$, proposed by \citet{Jones2009}, a solution with commonly assumed parameters is obtained, for example, for a Doppler factor of $\delta = 30$, a magnetic field of $B = 0.01$ G and a radius of $R \sim 1.6\e{17}$ cm, which is close to the limit set by the variability time-scale. A reduction in the size of the emission region would require an increase of the Doppler factor. The particle energy distribution is described with $\gamma_b = 4\e{4}$. The index variation between the first and second slopes of the BPL does not account for a simple synchrotron cooling break, due to the need of a relatively steep slope to match the \textit{Swift} XRT data. This points to the known limitation of the simple one-zone model, where acceleration, energy loss and particle escape are not explicitly modelled \citep{Katarzynski:2001aa}. In this scenario, the emitting region is relatively far from equipartition with a value of the electron energy density to magnetic energy density ratio, $u_e/u_B \sim 34$. However, such deviations are not unexpected for a source of type HSP like \src \citep[see e.g.][]{cerruti2013}. The parameter values of the SSC model are not different from the ones usually obtained for the other $\gamma$-ray HSP sources \citep[see e.g.][]{Tavecchio2010,Zhang2012}, showing that \src fits within the current population of known TeV blazars.

Considering that the $z = 0.17$ redshift from \citet{Jones2009} is uncertain (see Section~\ref{sec:Intro}),
it was decided to investigate whether the redshift $z = 0.06$ would yield solutions with more moderate model parameters,
in terms of $\delta$ or equipartition factor.
For an exemplary solution with small bulk Doppler factor, a value of $\delta = 20$ was chosen,
leading to a larger value of the magnetic field strength compared to the high-redshift solution.
The chosen set of parameters yields an electron and magnetic energy density ratio of $u_e/u_B \sim 23$, a bit closer to equipartition than for the solution at higher redshift. 

\begin{table*}
    \centering                          % used for centering table
    \begin{tabular}{ |c || c | c | | c | c ||}        % centered columns (4 columns)
       
        \cline{2-5}
        \multicolumn{1}{  c ||}{}    & \multicolumn{2}{c||}{z = 0.06}            & \multicolumn{2}{c||}{z = 0.17} \\   \cline{2-5}          % inserts double horizontal lines
        \multicolumn{1}{  c ||}{}    &  Range   &    Example                        &   Range       & Example \\
       
        \hline                                                                        % inserts single horizontal line
        $\delta$                &      [15, 52]                 & 20          &  [22, 100]              &  30 \\
        $K$ [1/cm$^3$]          &      [1.2\e{-7}, 3.7\e{-6}]             & 1.5\e{-7}           &  [0.2\e{-7}, 4.0\e{-6}]   &  0.3\e{-7} \\
        $R$ [cm]                &      [1.1\e{16}, 7.3\e{16}]   & 6.7\e{16}   &  [1.2\e{16}, 1.6\e{17}] &  1.6\e{17}\\
        $B$ [mG]                 &      [13, 49]           & 20        &  [3, 37]         &  10 \\
        $n1$                    &      Fixed                    & 1.7         &  Fixed                  & 1.7 \\
        $n2$                    &      Fixed                    & 3.5         &  Fixed                  & 3.5 \\
        $\gamma_{min}$           &      Fixed                    & 100         &  Fixed                  & 100  \\
        $\gamma_{b}$             &     [1.8\e{4}, 3.2\e{4}]       & 3\e{4}      &  [1.6\e{4}, 4.5\e{4}]     & 4\e{4} \\
        $\gamma_{max}$           &      Fixed                    & 5\e{6}      &  Fixed                  & 5\e{6} \\
        \hline
        \hline
        $u_e/u_b$                &      [5, 156]               & 23        & [8, 536]             &  34 \\
%        $L$ [erg/s]             &    [15\e{42}, 39\e{42}]        &     34\e{42}    &    [62\e{42}, 334\e{42}] &  176\e{42}  \\
        $L$ [$10^{43}$ erg/s]             &    [1.5, 3.9]        &     3.4    &    [6.2, 33.4] &  17.6  \\
        \hline
    \end{tabular}
    \caption[SSC model parameters for \src]{SSC model parameters for $z=0.06$ and $z=0.17$.
      The values in the "Range" column correspond to the allowed intervals obtained from the scan of the parameters space using the algorithm developed by \citet{cerruti2013}. The values in the "Example" colums correspond to particular solutions selected for illustration purpose.
      See text for the definition of the different parameters.}

    \label{table:modelparams}      % is used to refer this table in the text
\end{table*}

We note that the overall model does not account for the low-energy non-simultaneous radio data, which can in turn be ascribed to different larger regions of the jet. From the parameter ranges it can be seen that the SSC solutions for the lower redshift assumption ($z = 0.06$) are concentrated in a narrower domain in parameter space than the solutions for $z=0.17$. For the lower redshift, solutions can be found closer to equipartition and with more modest values of the bulk Doppler factor. Apart from these indications, no preference can be given to one or the other of the redshift estimates, based on the SSC model. It should be clear, however, that the scenario will require more extreme values if one assumed an even higher redshift for this source.

\section{Conclusions}
\label{sec:Conclusion}

We report the discovery with the H.E.S.S.\@ telescopes of VHE \g-ray emission from the HSP \src. The source was detected at $6\sigma$ level in 22.3 hours (live-time) with an average VHE \g-ray spectrum well described with a power law with a photon index $\Gamma=3.40\pm0.66_{\text{stat}}\pm0.20_{\text{sys}}$ and an integral flux above $\SI{200}{\GeV}$ corresponding to 1.1$\%$ of the Crab nebula flux. We report also the analysis of multi-wavelength data obtained at different times with \textit{Swift} UVOT \& XRT, \textit{RXTE} PCA, \fermilat, and additional data from \textit{WISE}, GROND, Catalina and ATOM. \textit{Swift} observed the source in different states of activity. For the state corresponding to the higher \textit{Swift} XRT flux, the source was quasi-simultaneously observed in the optical regime with GROND. These observations, along with contemporaneous infrared \textit{WISE} data and the $\sim1-2$ days variability observed by \textit{RXTE}, provide strong constraints for the description of the emission of the source in terms of synchrotron radiation.
Using the whole H.E.S.S.\@ data as an indicator of the source behaviour in the VHE \g-ray regime,
together with \fermilat data around the \textit{Swift} high state,
and considering two possible values $z = 0.17$ and $z = 0.06$ for the redshift of the source, we showed that a simple one-zone leptonic SSC model provides a good description of the broad-band emission of \src, with parameters compatible with the ones usually obtained for the other known TeV blazars.
The lack of a firm redshift is however an issue for the understanding of the source. In the absence of detection of spectral lines in the optical regime, constraints on the redshift could be provided by deeper TeV observations resulting in a significant detection at energies at or above 1 TeV, where the effects of EBL absorption will significantly differ between redshift assumptions \citep[see, e.g.][]{Mazin2007}.

\section*{Acknowledgements}
The support of the Namibian authorities and of the University of Namibia in facilitating the construction and operation of H.E.S.S. is gratefully acknowledged, as is the support by the German Ministry for Education and Research (BMBF), the Max Planck Society, the German Research Foundation (DFG), the Alexander von Humboldt Foundation, the Deutsche Forschungsgemeinschaft, the French Ministry for Research, the CNRS-IN2P3 and the Astroparticle Interdisciplinary Programme of the CNRS, the U.K. Science and Technology Facilities Council (STFC), the IPNP of the Charles University, the Czech Science Foundation, the Polish National Science Centre, the South African Department of Science and Technology and National Research Foundation, the University of Namibia, the National Commission on Research, Science \& Technology of Namibia (NCRST), the Innsbruck University, the Austrian Science Fund (FWF), and the Austrian Federal Ministry for Science, Research and Economy, the University of Adelaide and the Australian Research Council, the Japan Society for the Promotion of Science and by the University of Amsterdam.
We appreciate the excellent work of the technical support staff in Berlin, Durham, Hamburg, Heidelberg, Palaiseau, Paris, Saclay, and in Namibia in the construction and operation of the equipment. This work benefited from services provided by the H.E.S.S. Virtual Organisation, supported by the national resource providers of the EGI Federation.\\
This research has made use of the SIMBAD database,
operated at CDS, Strasbourg, France.\\
This publication makes use of data products from the Two Micron All Sky Survey, which is a joint project of the University of Massachusetts and the Infrared Processing and Analysis Center/California Institute of Technology, funded by the National Aeronautics and Space Administration and the National Science Foundation.\\
This research has made use of the NASA/ IPAC Infrared Science Archive, which is operated by the Jet Propulsion Laboratory, California Institute of Technology, under contract with the National Aeronautics and Space Administration. \\
The CSS survey is funded by the National Aeronautics and Space
Administration under Grant No. NNG05GF22G issued through the Science
Mission Directorate Near-Earth Objects Observations Program.  The CRTS
survey is supported by the U.S.~National Science Foundation under
grants AST-0909182 and AST-1313422.\\
This research has made use of data and/or software provided by the High Energy Astrophysics Science Archive Research Center (HEASARC), which is a service of the Astrophysics Science Division at NASA/GSFC and the High Energy Astrophysics Division of the Smithsonian Astrophysical Observatory.\\
Co-author M. Arrieta is supported by the Paris Science et Lettres (PSL) foundation. \\

%%%%%%%%%%%%%%%%%%%%%%%%%%%%%%%%%%%%%%%%%%%%%%%%%%



\bibliographystyle{mnras}
\bibliography{biblio} % if your bibtex file is called example.bib

%%%%%%%%%%%%%%%%%%%%%%%%%%%%%%%%%%%%%%%%%%%%%%%%%%

%%%%%%%%%%%%%%%%% APPENDICES %%%%%%%%%%%%%%%%%%%%%

%\appendix


%%%%%%%%%%%%%%%%%%%%%%%%%%%%%%%%%%%%%%%%%%%%%%%%%%

% Don't change these lines
\bsp	% typesetting comment
\label{lastpage}
\end{document}